# Influence of fillets onto mechanical properties of octet-truss lattice structures

Pierre-Thomas Doutre [a], Christelle Grandvallet [a], Léa Gobet [a], Frederic Vignat [a], Rémy Dendievel [b]

[a]Univ. Grenoble Alpes, CNRS, Grenoble INP, G-SCOP, Grenoble, F-38000, France.
[b]Univ. Grenoble Alpes, CNRS UMR 5266, Grenoble INP, SIMAP, Grenoble, F-38000, France.
Tel.: +334 76 57 43 57

Abstract

*The development of Additive Manufacturing (AM) for the fabrication of metallic parts allows structures to be directly manufactured from 3D models. The Electron Beam Melting (EBM) technology is an example of AM technologies that enables the manufacturing of new designs and sophisticated geometries. The process is particularly well suited for the fabrication of lattice structures. Octet-truss lattice structure has been a subject for research during the past 10 years. The potentials that it possesses attract enough interest for manufacturers to use it during the production of parts. Besides being lightweighted, the structure could provide solid mechanical properties. However, researchers always encounter the same issue regarding this particular structure. During Finite Element Analysis (FEA) simulation, stress concentration tends to appear at the struts intersection. This is due to the sharp edges that have very small surface area, thus provoking the presence of singularities. In this respect, the proposed solution is to integrate rounded-joints or fillets at the struts intersection. However, adding fillet entails a mass increase of octet-truss structures. To avoid this mass increase related to these fillets, it is necessary to reduce the size of octet-truss struts. This research work studies the influence of fillets onto the mechanical properties of structures with identical mass. To do so, a set of 15 octet-truss structures are designed with various fillet sizes and strut sizes and compared. Whereas some of them have thick struts and small fillets, others have smaller struts and bigger fillets. The main technical issue in this study remains the design of fillets for octet-truss structures. These latter can indeed be created for up to 12 struts that converge to the same point. Once designed, these octet-truss structures are fabricated by EBM technology and undergo static compression testing. Mechanical properties of each structure are finally determined. Results show that for the same relative density, octet-truss with fillets degrades the mechanical characteristics of the whole structures. This study shows that the strength/mass ratio is better for a structure without fillets. this result can be used in lightweighted applications.*

*Email address:* Pierre-thomas.doutre@grenoble-inp.fr (Pierre-Thomas Doutre)

# 1. Introduction, state of the art and objectives.

## 1.1. Introduction

The use of additive manufacturing technologies such as Electron Beam Melting enables the fabrication of complex geometrical entities, especially lattice structures [1]. The latter allows to reduce significantly the mass of products [1]. Comparative studies [2, 3, 4] have been led onto lattice structures in order to determine this influence and highlighted two types of structures:

- Bending dominated structures that have low rigidity, low resistance and high energy absorption capacity
- Stretching dominated structures that are more rigid and resistant, but have a lower energy absorption capacity [2, 3, 4 ,5].

Among these structures, octet-truss present a behaviour mainly dominated by stretching [6], which provides a high rigidity. For this reason, many research works have focused on ways to improve its performances. For instance, [7, 8, 9] studied octet-truss globally but also locally, revealing stress concentrations at nodes level. To overcome this issue, fillets can be added at nodes level, however this entails a mass increase of octet-truss structure. To avoid it, one solution is to reduce the size of octet-truss struts.

This paper therefore tackles the addition of fillets onto octet-truss structures. In order to maximize the strength/mass ratio, is it more relevant to populate a given volume with octet-truss structures having large struts without fillets or having large fillets with small struts? This raises the question of whether it is more appropriate to add material in the nodes or in the struts? First of all, a state of the art centered around the contribution of fillets is presented. It is followed by the design methodology and the experimental protocol about the addition of fillets. The obtained results are then studied, before concluding and evoking the limits of our study.

## 1.2. State of the art

In mechanical design, fillets enable smooth transitions between two surfaces generated by a sphere scanning. Several authors have tried to add fillets to demonstrate their benefits in terms of resistance increase. [10,11]. Yet, to add fillets increases the mass, therefore the density and resistance [2]. Other studies about the impact of fillets on the mechanical strength in fatigue have been conducted: [12] and [13] show that adding fillets avoids failures near the nodes [11] or at junctions with dense parts [14].

Arshad et al. [10] show by uniaxial compression tests that the addition of fillets onto octet-truss structures leads to an improvement of the mechanical characteristics (stiffness, energy absorption efficiency, tensile strength). The study is made at constant mass and volume. These structures have been built in PA12 with HP MJF4200 technology. Latture et al. [11] studied the impact of fillets in octet-truss structures. They compared a structure with fillets to one without fillets. The material used in this study is polycarbonate obtained by stereolithography. The relative density of the structures increases from 8.2% to 9.0% for a fillet radius of 1 mm. This study shows that the fillets used have a minor role in attenuating stress concentrations but allows to increase the nodal rigidity, and to reduce the degree of softening due to bending and increased the peak stress by 20%. The authors also show that there is a boundary effect on the mechanical property of lattice structures. They insist on the importance of the number of unit cell repetitions to limit these effects. Portela et al. [15] studied the impact of node topology on stiffness for different types of lattice structures (including the octet-truss). 6 types of nodes with fillets are studied in lattice structure. These lattices have a 5x5x5 tessellation of 7.2 mm unit cells, with strut diameters ranging from 0.55mm to 1.6mm. The relative densities of octet-truss samples ranged from 5.8% to 22.6%. The parts are manufacturing in polymer resin with DLP (digital light processing). They performed afterwards uniaxial compression. The authors observe an increase of 17 to 28% in stiffness when the nodal geometry is modified. The impact of nodal geometry on effective strength is not studied here.

In the previously mentioned articles, [11] and [15] evoke a gain of performance due to the presence of fillets. But this gain in performance does not occur at constant mass. On the other hand, the design approaches followed for building fillets unfortunately lacks of detail. Those which provide details are based on the use of fillets functions offered by CAD software support. That is why, based on these gaps, this article proposes to deepen the topic of fillet influence onto octet-truss and check the mechanical properties of such structures.

### 1.3. Objectives and research question

This article aims at finding the balance between fillet addition and reduction of strut radius in order to optimize the mechanical properties of octet-truss structures with a constant mass. We note that for an identical mass, the more important the fillets parameter, the smaller strut radius. The research question that arises is the following:

(i) At identical structure relative density, would fillets improve the mechanical strength of octet-truss structures for static load considering that it will involve strut diameter reduction?

(ii) What simulation tools should be used to design lattice structures with or without fillets?

## 2. Design and fabrication

Several octet-truss structures with different fillet sizes are designed in this study. This design phase is particularly complex as 12 struts have to meet at each node. The Rhino3D software Grasshopper module is used to generate the structures. In this module, we used the multipipe tool (Figure 1). In this tool, we used three inputs to generate the structures with fillets. The first input, "curves", corresponds to the lines or curves between which we want to generate fillets. The second entry, "node size", corresponds to Struts radius. The last one, "EndOffset", allows to control different fillet sizes. In the rest of this article, we will call this parameter « fillets ratio ». This parameter illustrated in Figure 2 is a strut radius multiplier. Point A corresponds to the intersection of the "struts" without fillets. Points B and B' correspond to the junction between the fillets and the struts. Fillets ratio (Fr) is defined as follows:

$$Fr = \frac{||\overrightarrow{AB}||}{R}$$

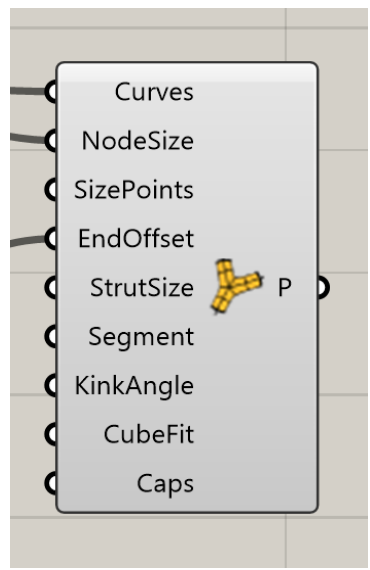

Figure 1: multipipe tools in rhino Grasshopper

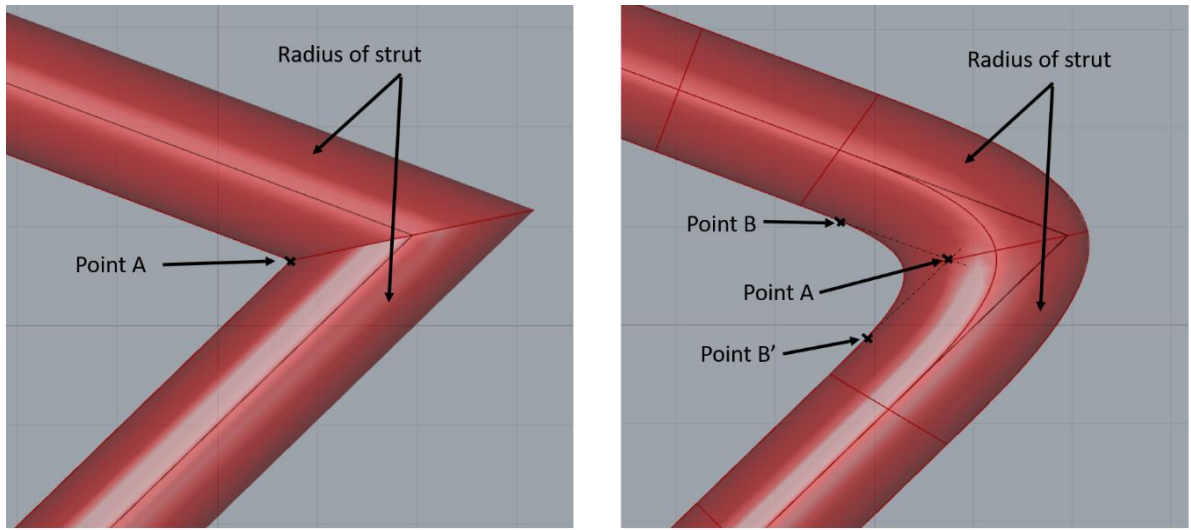

*Figure2: parameter fillets ratio structure without fillet (left) and structure with fillets (Right)*

Five structures of identical mass have then been designed. Figure 3 shows possible strut radius and fillets ratio for structures of identical relative density (3%). Each of them consisted in a repetition of 3x3x3 unit cell with a length of 20mm unit cells. This size is large enough to get reliable results in a reasonable manufacturing time. Figure 4 presents the structure with maximum value of fillets ratio (strut radius = 0.42 mm et Fillets ratio = 8.58mm/mm). The left figure shows a view of the complete structure, whereas the right figure is an enlargement of the complete structure. Table 1 represents all the parameters of the designed structures.

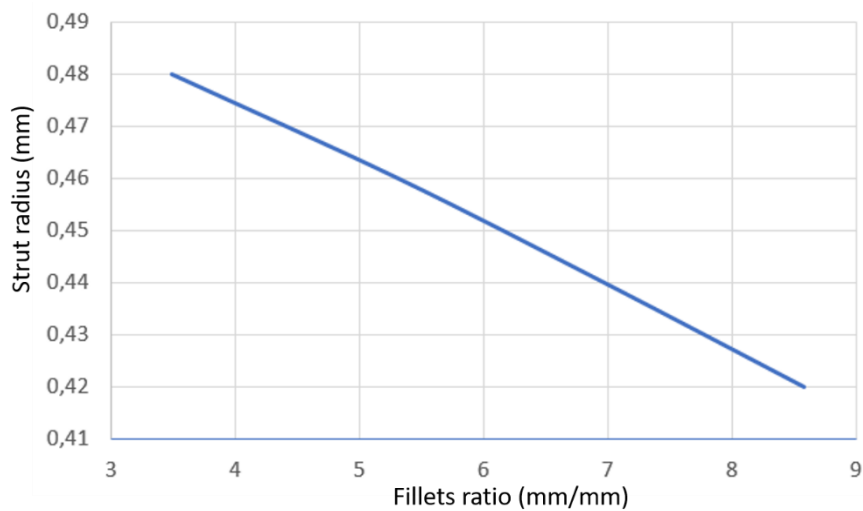

Figure 3: Parameters strut radius and fillets ratio for a structure octet-truss of same relative density (3%)

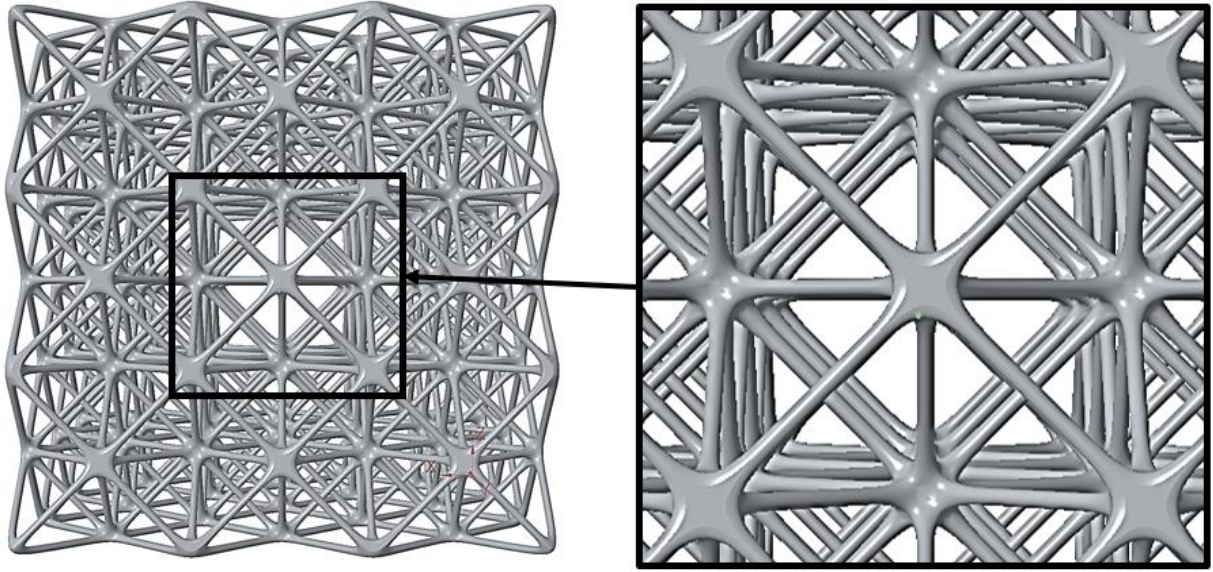

*Figure 4: Overall structure designed with maximum fillet (radius = 0.42mm, fillets ratio = 8.58 mm/mm)*

*Tab. 1: Summary of structure characteristics once designed*

| Structure numbers | Strut radius | fillets ratio | Volume | Relative density |
|---|---|---|---|---|
| (-) | (mm) | (mm/mm) | (mm$^3$) | (%) |
| 1 | 0,50 | - | 7845 | 3,63 |
| 2 | 0,48 | 3,485 | 7845 | 3,63 |
| 3 | 0,46 | 5,310 | 7845 | 3,63 |
| 4 | 0,44 | 6,970 | 7845 | 3,63 |
| 5 | 0,42 | 8,580 | 7845 | 3,63 |

The strut radius of the five structures has been purposely reduced in order to compensate for the gradual increase of fillets ratio and keep a constant mass.

As far as the manufacturing of these structures is concerned, the Electron Beam Melting (EBM) technology has been chosen, with TA6V alloy powder. EBM technology makes it possible to manufacture lattice structures [1], [16]. Past researches recommend a minimum radius of 0.4 mm [16] and provide a depowderable criteria [17] [18] in accordance with the manufacturability.

For each set of parameters, three samples are designed for a total of 15 structures to be fabricated.

Regarding their orientation, all 15 samples have a 45 degrees positioning into the build chamber

(Figure5). This orientation for lattice structures or thin struts is recommended in the works of [15], [19] and [16].

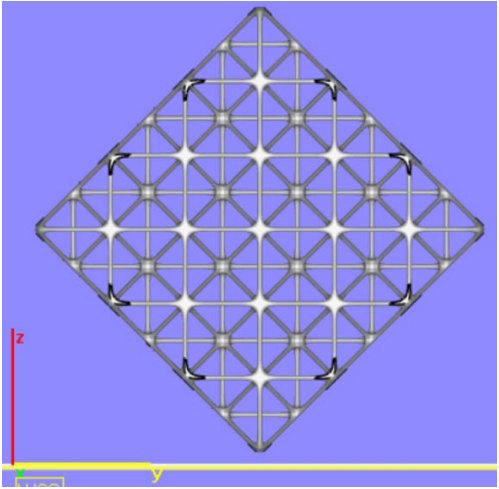

*Figure 5: Orientation of all 15 sample.*

The structures have been weighted at the end of the fabrication, in order to check the mass discrepancy. Table 2 summarizes the mass measurements carried out for each manufactured sample. The average mass of the samples is 34.4g and the gap from this average is shown in tab 2, maximum gap ranges from -6% to 4%.

*Tab. 2: Details of the 15 structures mass (in g)*

| Fillets ratio (mm/mm) | Strut radius (mm) | Sample mass (g) | Gap (%) |
|---|---|---|---|
| - | 0,5 | 29,7 | -5,45 |
|  |  | 29,9 | -4,82 |
|  |  | 29,9 | -4,82 |
| 3,485 | 0,48 | 30,9 | -1,63 |
|  |  | 30,9 | -1,63 |
|  |  | 31 | -1,32 |
| 5,31 | 0,46 | 32 | 1,87 |
|  |  | 31,7 | 0,91 |
|  |  | 31,7 | 0,91 |
| 6,97 | 0,44 | 32,2 | 2,50 |
|  |  | 32,1 | 2,19 |
|  |  | 32,6 | 3,78 |
| 8,58 | 0,42 | 32,1 | 2,19 |
|  |  | 32,1 | 2,19 |
|  |  | 32,4 | 3,14 |

Once fabricated, the 15 structures have been mechanically tested following an experimental protocol presented in the next section.

## 3. Experimental test and analyse method

A uniaxial compression has been implemented in order to get the mechanical properties of the 15 octet-truss structures. The tests are characterized by a sequence of loading-unloading. The same method as [1] has been applied for these tests. The compression bench is monitored in displacement with a speed of 1 mm per minute. Each 0.2% deformation, an unloading occurs up to 60% of the maximum charge. Results about the impact of fillets addition in octet-truss are detailed in the following section. The evolution of the stress-strain is plotted and shows in Figure 6.

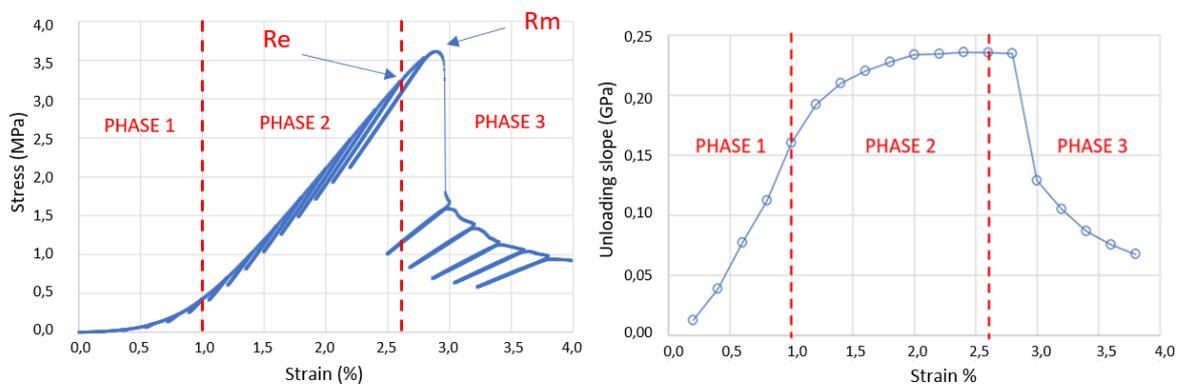

Fig. 6: Typical stress-strain cycle applied to a lattice structure for determining its characteristics Young's modulus (left). Unloading slopes as a function of the strain (right). Sample with strut radius 0.48 mm and fillets ratio of 3,485.

From both graphs figure 6, the compression of such structures can be divided into three phases:

**Phase 1** The first region corresponds to the positioning of the sample between the platens. If the opposite sides of the sample are not fully parallel and because roughness is present on the top and bottom surfaces, the first measurement of the force is not fully related to the elastic regime of the structure. The measured force is lower than the expected one. There is a rapid increase of the slope at unloading.

**Phase 2** After the first stage, the structure deforms in an elastic manner until the first strut becomes plastically loaded. The slope at unloading still increases in this region until it reaches a plateau. It means that at the early stages of elasticity, some struts are not fully loaded and do not act in the stiffness of the structure. In the plateau, the structure acts fully elastically.

**Phase 3** After a given strain, the first local damage occurs. It decreases the calculated unloading slope. The global structure reacts plastically.

The Young's modulus of the structure has been chosen as the maximum of the calculated unloading slopes (Figure 6 right). It corresponds to the value of the fully elastic plateau. In some structures,
the elastic plateau does not exist. Indeed, the structure is not fully elastically loaded when the first local plastic deformation occurs. Thus, the Young's modulus is taken as the maximum value. Re is defined as the beginning value of the stress at the phase 3. The tensile strength (Rm) corresponds to the maximal permissible stress in the structure (as illustrated on figure 6, left).

## 4. Results

### 4.1. Comparison with a literature result

Before discussing the mechanical tests results, their consistence is verified by comparing them to results from [8] and [20]. These papers proposes mechanical characteristics of octet-truss lattice structures without fillets. The manufactured structures have a relative density of 3%. For this relative density, without fillets and concerning the relative young modulus it can be noticed a difference of 12% less compared to the results of [8] and 15% less compared to the results of [20] (illustrated in tab 3). The results of the tests presented in this paper can thus be considered for further discussion.

*Tab. 3: comparison with literature results*

|  | Relative density (%) | Relative young's modulus (%) | Gap (%) |
|---|---|---|---|
| Deshpande and al. [8] | 3 | 0,25 | 12 |
| Azman [20] | 3 | 0,26 | 15 |
| Results of mechanical tests | 3 | 0,22 | - |

### 4.2. Results analysis

The overall results are presented on Figure 9. The graphs show all measurements of Re and Rm (bottom graph figure 9) and E (top graph figure 9) according to the strut radius. It may be noted that the evolution of the mechanical characteristics (Re, Rm and E) decreases while the fillet ratio increases.

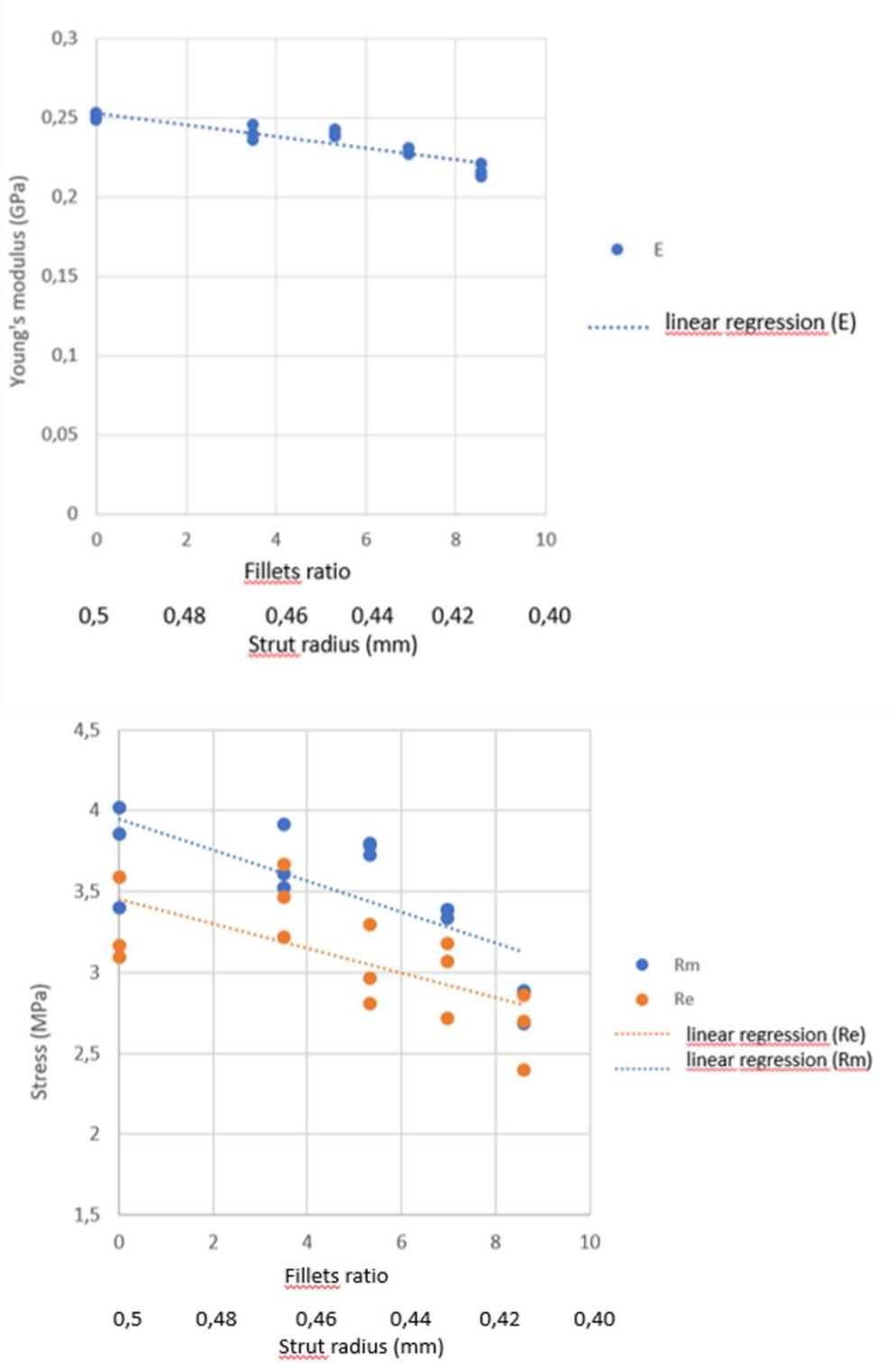

*Fig. 1 : Overall results of mechanical testing*

Note that, the deviation of the mass reported after the manufacturing process is not in contradiction with the previous results. Because if the structures were of constant mass, the difference in mechanical properties between the structures with and without fillets would be even greater.

This negative impact of fillets is verified by finite element calculations as shown in figure 10 where several volume finite element calculations have been performed. The calculation is performed on a unit cell to avoid cumbersome computation resource with complete volumetric structures. The boundary conditions on both structures are identical (a force of 100N is applied on the upper part and all degrees of freedom are blocked in the lower part). Figure 10 shows the distributing of Von Mises stress with their parameters strut radius and fillets ratio.

The first observation is that, in all structures, the stresses in the fillets are lower than in the struts. This means the elastic domain limit is reached in the strut before reached the fillets. This is consistent with the calculations and experimental results obtained by [21]. The second observation is that, for a given load and mass, higher fillets ratio values leads to lower strut diameters and as a consequence higher stress in the strut, which confirms the experimental results we have obtained.

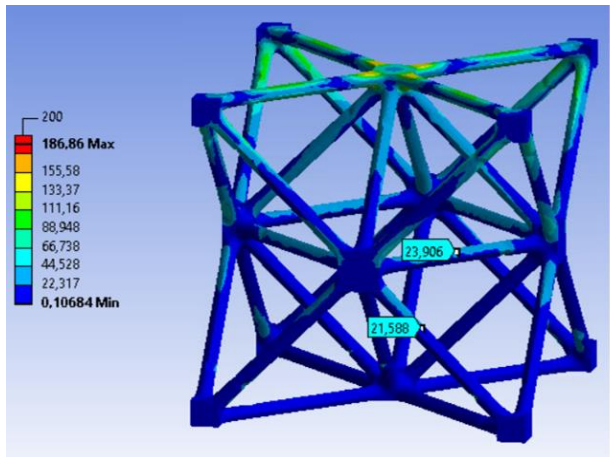
Strut radius 0,48mm and fillets ratio 3,485

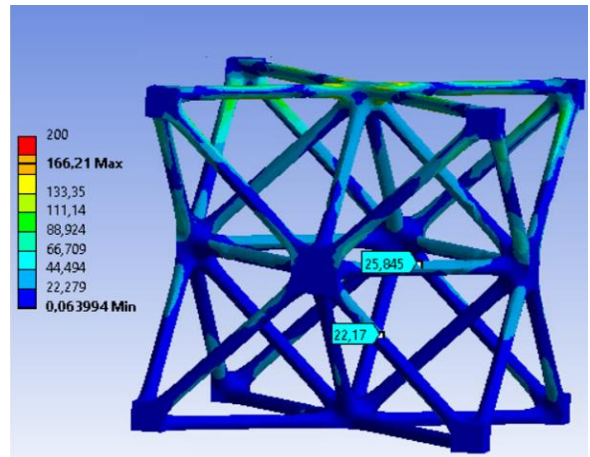
Strut radius 0,46mm and fillets ratio 5,310

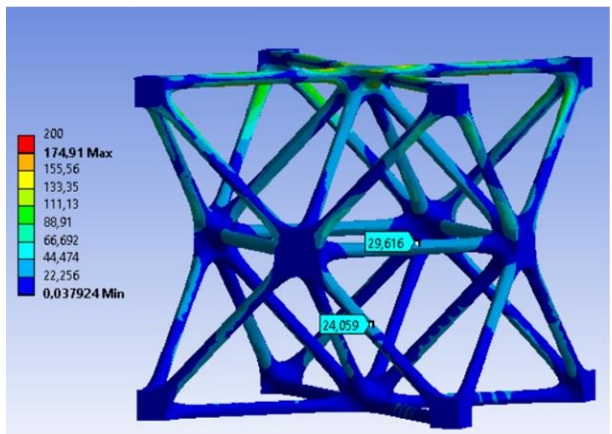
Strut radius 0,44mm and fillets ratio 6,970

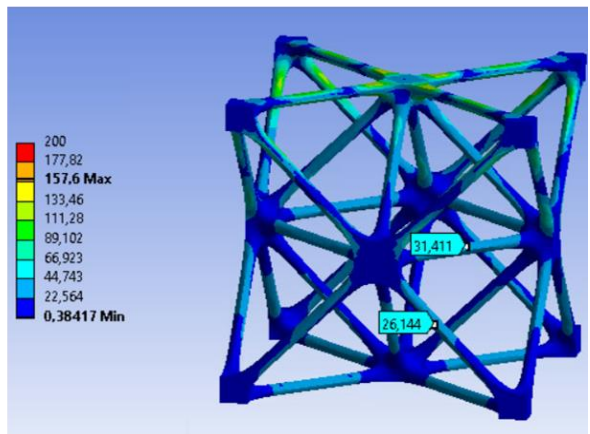
Strut radius 0,42mm and fillets ratio 8,580

*Figure 10: Von Mises stress calculation for 4 structures with different fillets ratio and strut radius values.*

4.3. Failure zone analysis:

The results obtained previously by simulation, can be verified experimentally. Indeed, the mechanical tests were filmed. The images of the failures are shown on figure 11 for each value of radius with a zoom on the right part.

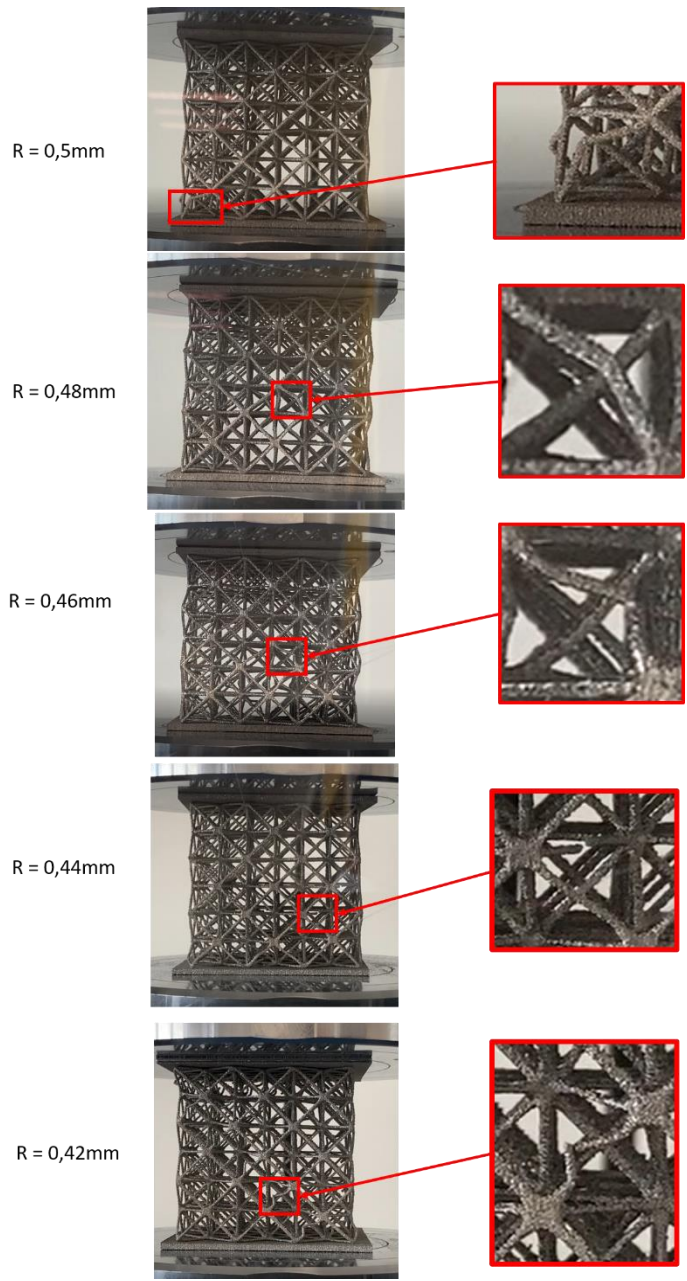

*Figure 11: representation of the failures*

It can be noted that for all mechanical tests, the failure appears systematically in the struts and not in the nodes or fillets. This finding is in agreement with the previous sections.

*4.4.* Use of beam model for lattice dimensioning

In sections 2.2 and 2.3, finite element simulations as well as mechanical tests shows that the failure of the structures occurred in the strut and not in the fillets. They also show that, to improve strength/mass ratio these structures should be designed without fillet. The question is then the following: is a beam model sufficient to dimension octet-truss structures? These calculations are valid for a density lower than 5%, which is the case here. As beam model is used, fillets are not taken into account which is consistent with the previous chapter conclusion. A beam based simulation has been performed for each value of the strut diameter. The applied boundary conditions are shown in figure 12. On each structure, a force along -y is applied on the set of nodes of the upper face and the displacement following y are blocked and rigid body motion is prevented by adequate boundary conditions. These boundary conditions allow to stay as close as possible to the above-described compression test. The simulations were performed in Ansys APDL. Table 4 summarizes the results of the simulations performed. It lists the value of the following parameters:

F1 is the experimental force for which the yield stress of the structure (Re) is reached. F2 is the same force determined using finite element simulation. It is thus the theoretical force for which the yield stress of the structure (Re) is reached (i.e. when the yield stress 840 MPA in a strut is reached). It has also been verified that the Euler buckling criterion is not reached in any of the tested configurations.

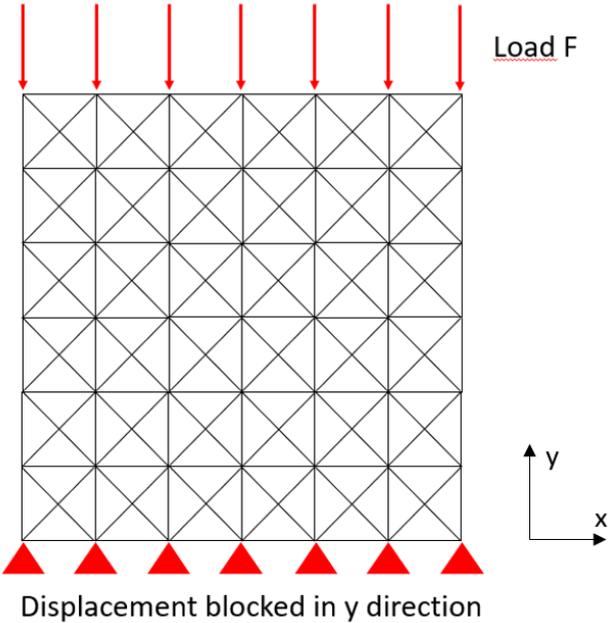

Figure 12: FEA modelling of complete lattice structure.

Table 4: FEM beam simulation data.

| Fillets ratio | Strut radius (mm) | F1 (N) | F2(N) | F2/F1 |
|---|---|---|---|---|
| - | 0,5 | 11844 | 17094 | 1,44 |
| 3,485 | 0,48 | 12420 | 16150 | 1,30 |
| 5,31 | 0,46 | 10908 | 14972 | 1,37 |
| 6,97 | 0,44 | 10764 | 13825 | 1,28 |
| 8,58 | 0,42 | 9540 | 12740 | 1,34 |

It can be observed that the ratio F2/F1 (theoretical force / experimental force) is stable and measure a difference 30% to 40% between the experimental force and the theoretical force. This difference can be explained by the concept of equivalent diameter introduced by [1] In this an equivalent diameter concept is proposed. This is the diameter of a strut modelling the material actually contributing to the mechanical strength. This diameter takes into account manufacturing deviation. For the machine used for the test, this diameter is smaller than the nominal one (CAD diameter) from 15% to 45% depending on the strut's orientation. The cross-section of the strut is then reduced by 27% to 70%. As the tested structure is stretching dominated, the force F1 is also reduced by 27 to 70% leading to a F2/F1 ratio between 1.37 and 3.

It can thus be concluded that the simulations carried out with beam models give results consistent with the experimental ones. These kinds of modelling can thus be considered when dimensioning lattice structures.

## 5. Conclusion

### 5.1. Discussion

The finite element simulations carried out with and without fillets show that fillets have a negative impact on the mechanical property for static loads. The experimentation of this research work also concludes that failures mainly appear at the struts for the structures with fillets as well as for the structures without fillets.

These observations are consistent with those reported by Latture et al [11] who observe a failure for structures with fillets that occurs at struts during static compression tests.

However, for structures without fillets, the results presented in our study are in contradiction with those of Latture et al who observe a break at nodes. Latture et al use another manufacturing process than ours (SLA) as well as another material (polycarbonate), which can explain this difference. However, although the resolution of their machine along the z axis is the same as ours (50 microns), the effect of smoothing during slicing cannot explain this difference.

On the other hand, comparing our results with authors who used the same metallic process and same material, such as Burr et al [21] who, in fatigue tests, observed failure in struts for titanium manufactured by EBM, our structures without fillets are similar. [21] also performed an X-ray tomography that showed that the defects in the nodes are as numerous as anywhere else in the structure.

### 5.2. Contribution

This research work focused on the influence of fillets onto mechanical properties of octet-truss lattice structures. To achieve it, five types of structures with different fillets values have been built in triplicate. Since fillets entail a mass increase onto the structures, it is decided to reduce the strut size so as to work with structures of identical mass. The results of static compression testing lead to the conclusion that, at identical mass, fillets do not allow to improve the mechanical characteristics of structures in statics, and even tend to degrade them. The analysis of the failure zones show that the plasticization and the failures occur in the struts and not in the fillets. In this case, during the design phases, we can consider the beam calculation is sufficient to design truss structures without fillets but for a low density (less than 5%). When designing octet-truss structures, the designer has little interest in adding fillets if the objective is to maximize the mass/resistance ratio. But if it is to perform finite element calculations on structures without fillets, this can be done on a beam model.

*5.3. Perspectives*

This experimentation was led on octet-truss structures with a strut size close to the limits of EBM process manufacturability. Researches about the impact of fillets onto mechanical properties of other structure types (for instance, cubic) could be of interest. For example, to study the impact of fillets with higher density, with loading cases in other directions, with a conical beam, with more unit cell repetition (4x4x4,5x5x5, etc …), with another material, with another manufacturing process or else the impact of fillets onto fatigue strength would be perspectives that could complete results about lattice structure behaviours at similar density.